\documentclass[aps,superscriptaddress,amsmath,amssymb,twocolumn,prl,showpacs]{revtex4}
\usepackage{hyperref}
\usepackage{epsfig}
\usepackage{subfigure}
\usepackage{braket}
\usepackage{bm}
\usepackage{color}

\newcommand{\rem}[1]{}

\newcommand{\beq}{\begin{equation}}
\newcommand{\eeq}{\end{equation}}
\newcommand{\beqa}{\begin{eqnarray}}
\newcommand{\eeqa}{\end{eqnarray}}

\newcommand{\refe}[1]{\eqref{#1}}
\newcommand{\refE}[1]{Eq.~\eqref{#1}}

\begin{document}
\title{Superharmonic long-range triplet current in a diffusive Josephson junction}

\author{Caroline Richard} 
\author{Manuel Houzet}
\author{Julia S. Meyer}
\affiliation{SPSMS, UMR-E 9001 CEA/UJF-Grenoble 1, INAC, Grenoble, F-38054, France}


\begin{abstract}
We study the Josephson current through a long ferromagnetic bilayer in the diffusive regime.
For non-collinear magnetizations, we find that the current-phase relation is dominated by its second harmonic, which corresponds to the long-range coherent propagation of two triplet pairs of electrons.
\end{abstract}

\date{\today}

\maketitle

The interplay between ferromagnetism and superconductivity in hybrid structures is an active area of research \cite{buzdin-rev,bergeret-rev}. In a homogeneous ferromagnet (F) adjacent to a conventional superconductor (S), superconducting correlations between pairs of electrons are induced both in the spin singlet channel and the triplet channel without spin projection along the magnetization axis. This proximity effect is short-ranged due to the dephasing of electrons with opposite spins induced by the ferromagnetic exchange field. On the other hand, in the presence of a non-collinear magnetic configuration, a long-range proximity effect can be induced in the triplet channels with parallel electron spins, as no such dephasing occurs in that case \cite{bergeret,kadigrobov}. 

Early proposals to observe this effect suggested measuring the critical current in a Josephson junction through a ferromagnetic {\it trilayer} with non-collinear magnetizations \cite{braude,houzet}. In this geometry, the external F layers convert singlet Cooper pairs into triplet pairs which have a non-vanishing projection onto the channels with parallel electron spins along the (tilted) magnetization of the central F layer, and thus may propagate coherently over long distances. A maximal critical current with amplitude comparable to that of a normal metallic Josephson junction with the same length is obtained when the external layers have a thickness comparable to the ferromagnetic coherence length, $\xi_F$, and the magnetizations in successive layers are orthogonal. (Similarly, long-range triplet correlations can be induced by spin-active interfaces between a ferromagnet and the superconducting leads \cite{eschrig,asano,eschrig2}.) Indeed, recent experiments observed a strong enhancement of the Josephson current through a ferromagnetic multilayer when the layers were ordered non-collinearly \cite{birge,robinson,meso}.

Here we are interested in the question whether three layers (or, equivalently, spin-active interfaces on both sides of the junction) are necessary to observe a long-range triplet Josephson current. Recently it has been shown theoretically that a long-range triplet proximity effect may also develop in ballistic {\it bilayer} ferromagnetic Josephson junctions  with non-collinear magnetizations \cite{trifunovic}. In this case, a superharmonic Josephson relation is generated by the long-range propagation of  an even number of triplet pairs which may then recombine into singlet Cooper pairs. It is important to know whether this effect is robust to disorder which is believed to be present in the experiments, i.e., whether it exists in diffusive systems as well. In this work, we show that this is indeed the case. We find that the amplitude of the critical current determined by the second harmonic of the Josephson relation decays algebraically with length. The maximal current is smaller than in trilayers due to fact  that the triplet pairs have to  recombine into singlet Cooper pairs, a process that takes place only within a distance $\sim\xi_F$ near the F/S interface. 

%
%
\begin{figure}\centering
\includegraphics[width=0.7\linewidth]{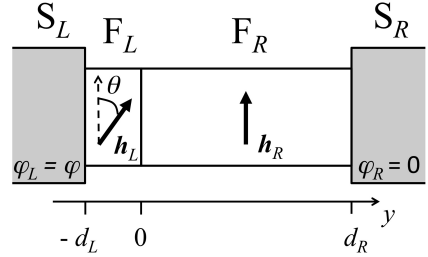}
\caption{
Setup of the junction. The superconducting leads are coupled through two ferromagnetic layers in series, with thicknesses $d_L$ and $d_R$, respectively. The magnetizations of the layers are tilted with respect to each other by an angle $\theta$.}
\label{junction}
\end{figure}

We consider a Josephson junction formed of two superconducting leads contacted through two ferromagnetic layers in series, with thicknesses $d_L$ and $d_R$, respectively, cf. Fig.~\ref{junction}. The F layers are assumed to have the same properties, but different orientations of their magnetizations. As we are interested in the long-range proximity effect, moreover, we assume that the right layer of length $d_R$ is much thicker than the diffusive ferromagnetic coherence length, $\xi_F=\sqrt{D/h}$, where $D$ is the diffusion constant and $h$ is the amplitude of the exchange field. 

Within the quasiclassical diffusive theory \cite{usadel}, the equilibrium supercurrent flowing through the junction can be expressed through  the quasiclassical Green function $g$, which is a $4\times 4$ matrix in the particle/hole and spin spaces, and obeys normalization conditions $g^2=1$ and $\mathrm{Tr}\;g=0$. It solves the nonlinear Usadel equation
\begin{equation}
-D\partial_y(g\partial_y g)
+\left[ (\omega  +i \bm{h}(y).\bm{\sigma})\tau_z, g\right]
=0,
\label{eq:Usadel}
\end{equation}
where $y$ is a coordinate along the junction, $\tau_i$ and $\sigma_j$ ($i, j =x,y,z$) are Pauli matrices in the particle/hole and spin spaces, respectively, and $\omega=(2n+1)\pi T$ ($n$ integer) is a Matsubara frequency at temperature $T$.  The orientations of the exchange fields are characterized by a tilt angle,
\begin{equation}
\bm{h}(y) =
h[\sin\theta(y) \;\bold{\hat x}+\cos\theta(y) \;\bold{\hat z}],
\end{equation}
where $\theta(y)=\theta$ for $-d_L<y<0$ and $\theta(y)=0$ for $0<y<d_R$.
Note that the orbital effect is neglected in \refE{eq:Usadel} by assuming that the magnetic flux through the junction is much smaller than the flux quantum. 

Equation \refe{eq:Usadel} has to be supplemented by boundary conditions at each interface. We assume that the interface resistance between the two F layers is much smaller than the resistance of each layer. Therefore, $g$ and its derivative are continuous at $y=0$. 
Moreover, assuming a rigid boundary condition at the interfaces with the leads as well as negligible resistances of the F/S interfaces, we impose the continuity conditions $g(-d_L)=g_L$ and $g(d_R)=g_R$, where $g_k=[\omega \tau_z+\Delta(\cos\varphi_k \tau_x-\sin\varphi_k\tau_y)]/\sqrt{\omega^2+\Delta^2}$ ($k=L,R$) are the Green functions of the leads. Here $\Delta$ is the amplitude of the superconducting gap and $\varphi_k$ its phase. For convenience, we choose $\varphi_L=\varphi$ and $\varphi_R=0$ such that $\varphi$ denotes the superconducting phase difference.

The supercurrent is then related to the Green function $g$ through
\begin{equation}
I= \pi e\nu D S T\sum_{\omega>0 } \mathrm{Im} \,\mathrm{Tr}[\tau_3  g\partial_y   g],
\label{eq:I}
\end{equation}
where $\nu$ is the density of states (per spin) at the Fermi level and $S$ is the cross section of the junction. Due to current conservation, \refE{eq:I} may be evaluated at any position along the junction.

To proceed further, we introduce the parametrization 
\begin{equation}
g=
\left(\begin{array}{cc}
\sqrt{1- F\tilde  F} & F\\
\tilde F  & -\sqrt{1- \tilde F F} 
\end{array}
\right),
\label{eq:g}
\end{equation}
which automatically satisfies the normalization conditions. Here, the anomalous functions $F$ and $\tilde F=\sigma_y F^* \sigma_y$ are $2\times 2$ matrices in spin space and odd in $\Delta$. With the parametrization \refe{eq:g}, the Usadel equation \eqref{eq:Usadel} takes the form
\beqa
-D\partial_y \left[ 
\sqrt{1- F\tilde F}\, \partial_yF
-F\,\partial_y\sqrt{1- \tilde F F}  
\right]
&&
\nonumber\\
+\left\{\omega  +i \bm{h}(y).\bm{\sigma}, F\right\}
&=&0,
\label{eq:Usadel2}
\eeqa
while the current \refe{eq:I} simplifies to
\begin{equation}
I= 2\pi e\nu D S T \sum_{\omega>0 }  \mathrm{Im}\,\mathrm{Tr}[F\partial_y\tilde F].
\label{eq:I2}
\end{equation}
Note that, as $F$ and $\tilde F$ are odd functions of $\Delta$, the current is  even in $\Delta$.

At temperatures slightly below the superconducting critical temperature $T_c$, the gap vanishes as $\Delta\propto\sqrt{T_c(T_c-T)}$. Thus, one may solve \refE{eq:Usadel2} perturbatively around the normal state solution, $F=0$.   To this end, we expand $F$ in the small parameter \cite{foot1}  $\Delta/\omega$, i.e., $F=(\Delta/\omega) F^{(1)}+(\Delta/\omega)^3 F^{(3)}+...$, and solve \refE{eq:Usadel2} order by order. To obtain the current up to the fourth order in $\Delta/T_c$, it is sufficient to compute $F^{(1)}$ and $F^{(3)}$.

Let us start with the leading order. Then, \refE{eq:Usadel2} yields the {linear} differential equation,
\begin{equation}
-D\partial^2_y F^{(1)}+2\omega F^{(1)} + i \left\{\bm{h}(y).\bm{\sigma}, F^{(1)}\right\}
=0.
\label{eq:F}
\end{equation}
Upon performing a unitary transformation, $ {\cal F}^{(1)}(y)=e^{i\sigma_y\theta(y)/2} F^{(1)}(y)e^{-i\sigma_y\theta(y)/2}$, the general solution in each layer takes the form ${\cal F}^{(1)}={\cal F}^{(1)}_0+{\cal F}^{(1)}_x\sigma_x+{\cal F}^{(1)}_z\sigma_z$, where 
\beq
\label{eq:FR}
{\cal F}^{(1)}_{L/R,s}=
A_s^{L/R} e^{-p_s y}+B_s^{L/R}  e^{p_s y}
\eeq
with $s=\pm,x$ and ${\cal F}^{(1)}_{\pm}={\cal F}^{(1)}_0\pm {\cal F}^{(1)}_z$. Here ${\cal F}^{(1)}_{\pm}$ with $p_\pm=\sqrt{2(\omega\pm i h)/D}$ correspond to the short-range singlet and triplet correlations, while ${\cal F}^{(1)}_x$ with $p_x=\sqrt{2\omega/D}$ corresponds to an equal superposition of the long-range triplet correlations. For typical ferromagnets, $h\gg T_c$, therefore, $p_\pm\approx (1\pm i)/\xi_F$. 

The coefficients $A_s^{L/R}$ and $B_s^{L/R}$ that enter Eqs.~(\ref{eq:FR}) are determined by the boundary conditions. Assuming $d_R\gg\xi_F$, we find $A_\pm^R=e^{i\varphi}a_\pm$, $B_\pm^R =e^{-p_\pm d_R}$, and $A_x^Re^{-p_xd_R}=-B_x^Re^{p_xd_R}=ie^{i\varphi}a_x$ (up to exponentially small corrections in $d_R/\xi_F$) with 
\begin{subequations}
\label{eq:coefs}
\beqa
a_+=a_-^*&=&\frac{\alpha-\beta}{|\alpha|^2-|\beta|^2},
\\
a_x&=&\frac{\tan\theta}{2\sinh p_xd}\mathrm{Im}\left[(p_+d_L+1)a_+\right],
\eeqa
\end{subequations}
and
\begin{subequations}
\label{eq:coefs2}
\beqa
\alpha&=&\cos^2\frac\theta 2 e^{p_- d_L}
\nonumber\\
&&
+\frac{\sin^2\theta}{2\cos\theta}(p_- d_L+1)\frac{\cosh p_- d_L \sinh p_x d_R}{\sinh p_xd}, 
\quad \\
\beta&=&\sin^2\frac\theta 2 (\cosh p_+ d_L- i \sinh p_+ d_L)
\nonumber \\
&&-\frac{\sin^2\theta}{2\cos\theta}(p_- d_L+1)\frac{\cosh p_+ d_L\sinh p_x d_R}{\sinh p_xd}, 
\quad
\eeqa
\end{subequations}
where $d=d_L+d_R$ is the total length of the junction.  We do not give the explicit results in the left layer because, in the following, we choose to evaluate the current \eqref{eq:I2} in the right layer.

Note that the ratio $d_L/\xi_F$ may be arbitrary, while Eqs.~\refe{eq:coefs} and \refe{eq:coefs2} have been simplified with the assumption $d_L\ll\xi_N$, where $\xi_N=\sqrt{D/(2\pi T_c)}$ ($\xi_N\gg \xi_F$) is the normal coherence length close to $T_c$. For a non-collinear configuration of the layers, long-range correlations are present (${\cal F}^{(1)}_x\neq0$). Nevertheless, they do not contribute to the first harmonic of the current-phase relation, as we show now.

Inserting the solution \refe{eq:FR} for ${\cal F}_R^{(1)}$ into \refE{eq:I2}, we obtain the first harmonic $I_1$ of the current-phase relation, $I_1(\varphi)=I_1\sin\varphi$, where $I_1\propto{\rm Re}[p_+e^{-p_+d_R}a_+]$. It takes the form
\beq
I_1= \frac{\pi  G\Delta^2}{\sqrt{2}e T_c}\frac{d}{\xi_F}e^{-d/\xi_F}\Upsilon_1\left(\frac{d_L}{\xi_F},\frac{d_R}{\xi_F},\theta\right),
\label{eq:I0}
\eeq
where $G=2e^2\nu D S/d$ is the conductance of the junction in the normal state and $\Upsilon_1$ is a scaling function.

\begin{figure}
\centering
\includegraphics[width=0.9\linewidth]{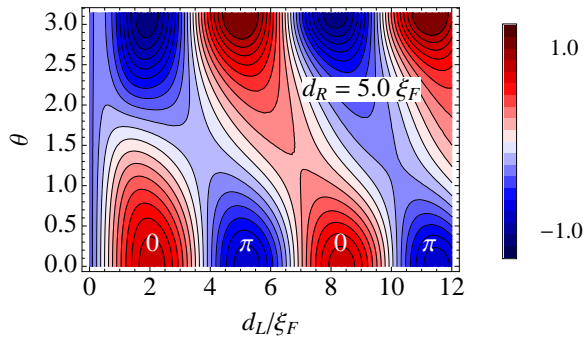}
\caption{
Dependence of the short-range first harmonic $I_1$ of the current-phase relation on the thickness $d_L/\xi_F$ and the angle $\theta$ at fixed $d_R/\xi_F=5$. Here $I_1$ is measured in units of
$I_{10}=\pi G\Delta^2 d/(\sqrt{2}e T_c\xi_F)e^{-d/\xi_F}$.}
\label{1st-harmonic}
\end{figure}

The exponential reduction factor $e^{-d/\xi_F}$ in \refE{eq:I0} shows that $I_1$ is always short-ranged. The scaling function $\Upsilon_1$  is shown in Fig. \ref{1st-harmonic} as a function of the length of the left ferromagnet and the tilt angle between the magnetizations. Its maximal absolute value is of the order of 1 and its sign oscillates, thereby displaying transitions between $0$-states (when $I_1>0$) and $\pi$-states (when $I_1<0$). Such oscillations have indeed been observed in S/F/S junctions \cite{ryazanov,kontos,foot2}. 

Simple analytic expressions for the function $\Upsilon_1$ in \refE{eq:I0} can be obtained in various regimes. For a parallel alignment, $\theta=0$, the results for a long ($d\gg\xi_F$) monodomain S/F/S junction are reproduced with $\Upsilon_1\simeq\sin(d/\xi_F+\pi/4)$ \cite{buzdin-pi}.  Moreover, as long as $d_L\ll\xi_F$, this result remains valid for arbitrary angles. In the opposite limit, $d_L\gg \xi_F$, we find $\Upsilon_1\simeq\sqrt{2}\cos(d_L/\xi_F)\cos(d_R/\xi_F)$ for $\delta\theta<\theta <\pi-\delta\theta$ and $\Upsilon_1\simeq \sqrt{2}\cos[(d_L-d_R)/\xi_F]$ for an antiparallel alignment, $\theta=\pi$. Here $\delta \theta \sim \xi_F/\mathrm{min}(d_L,d_R)$ is the crossover scale over which $\Upsilon_1$ varies  smoothly between its respective values at $\theta=0$, at intermediate angles, and $\pi$. These results generalize those of Refs.~\cite{blanter,crouzy} obtained at $d_L=d_R$ \cite{foot3}.

We now turn to higher-order contributions in $\Delta/T_c$ to the current-phase relation. The fourth-order terms consist in a small (short-range) correction to the amplitude $I_1$ of the first harmonic, as well as a second harmonic $\propto\sin(2\varphi)$. Here we concentrate on the second harmonic. In collinear structures, its amplitude $I_2$ is short-ranged with a  suppression factor $e^{-2d/\xi_F}$. Thus, it becomes important only near $0/\pi$-transitions \cite{buzdin-superharm,houzet-superharm} when $I_1$ vanishes. In non-collinear structures, however, $I_2$ is long-ranged as we show now.

\begin{figure}
\centering
\textbf{a)}\hspace*{0.98\linewidth} 
\\
\includegraphics[width=0.9\linewidth]{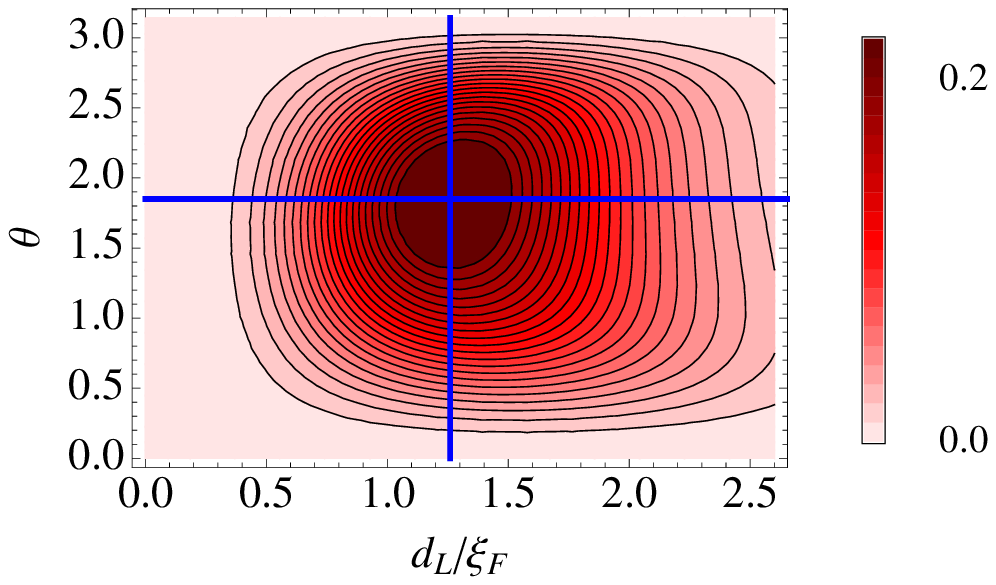}
\\
\textbf{b)}\hspace*{0.465\linewidth} \textbf{c)}\hspace*{0.515\linewidth} 
\\
\includegraphics[width=0.49\linewidth]{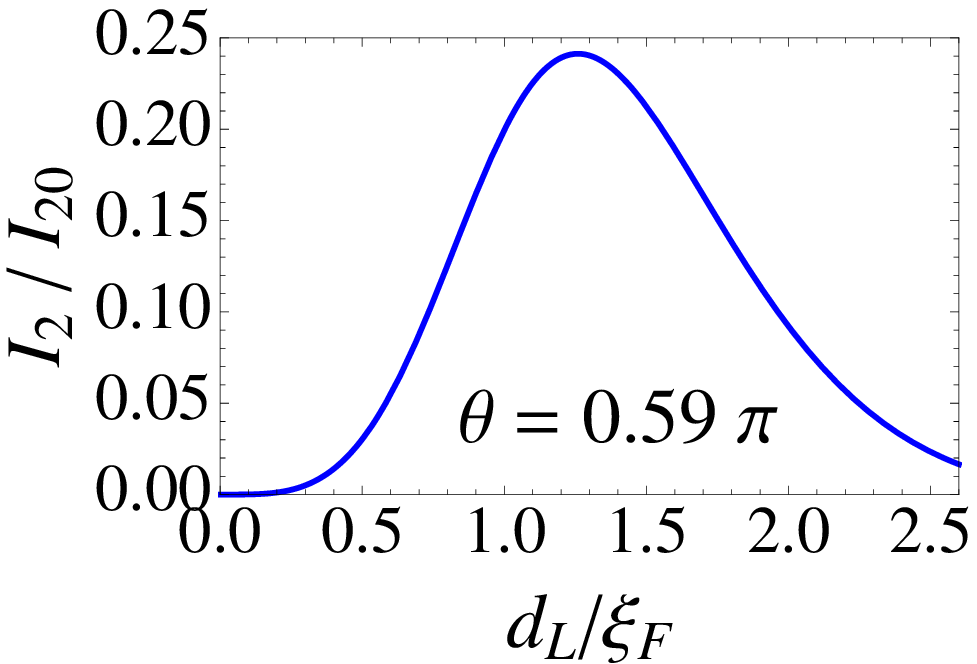}\hfill
\includegraphics[width=0.49\linewidth]{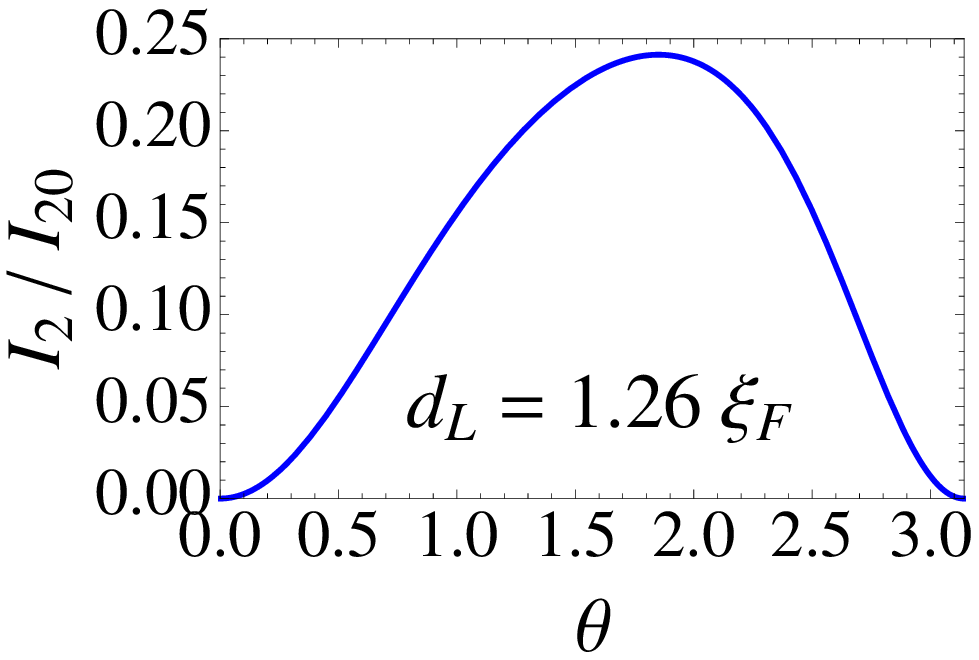}
\caption{
Dependence of the long-range second harmonic $I_2$ of the current-phase relation on the thickness $d_L/\xi_F$ and the angle $\theta$. Here $I_2$ is measured in units of
$I_{20}=\pi G\Delta^4 \xi_F/(192 e T_c^3d_R)$.
The plots b) and c) are taken along the cuts indicated by the blue lines in a),
at $\theta=0.59\pi$ and $d_L/\xi_F=1.26$, respectively. Both cuts include the maximal value $I_2^{\rm max}\approx0.24I_{20}$.
}
\label{2nd-harmonic}
\end{figure}

For concreteness, we will assume $d_R\gg d_L$ and evaluate the current \refe{eq:I2} in the right layer. Long-range contributions originate from the propagation of parallel-spin triplet pairs described by $F_x$, namely $I_2^{lr}\propto \mathrm{Im} \,\mathrm{Tr}[F_{R,x}^{(1)}\partial_y\tilde F_{R,x}^{(3)}+F_{R,x}^{(3)}\partial_y\tilde F_{R,x}^{(1)}]$. Using \refE{eq:Usadel2}, we find that $F_{R,x}^{(3)}$ obeys the equation
\begin{widetext}
\beqa
-D\partial_y^2 F_{R,x}^{(3)}
+2\omega F_{R,x}^{(3)}
=
\frac D 2\partial_y
\left\{
\left[
(F_{R,x}^{(1)})^2+F_{R,+}^{(1)}F_{R,-}^{(1)}
\right]
\partial_y \tilde F_{R,x}^{(1)}
+
\left[
F_{R,+}^{(1)}\partial_y \tilde F_{R,+}^{(1)}
+
F_{R,-}^{(1)}\partial_y \tilde F_{R,-}^{(1)}\right]
F_{R,x}^{(1)}
\right\}.
\label{eq:Usadel3}
\eeqa
\end{widetext}
Being interested in the second harmonic, we only need to retain contributions to $F_{R,x}^{(3)}$ that are proportional to $e^{-i\varphi}$. Thus, we may decompose $F_{R,x}^{(3)}=f_x^{(3)} e^{-i\varphi}+\Phi_{R,x}^{(3)}$, where $\Phi_{R,x}^{(3)}$ does not contain any contributions $\propto e^{-i\varphi}$. The function $f_x^{(3)}$ then obeys a differential equation obtained from  \refE{eq:Usadel3} by keeping on the r.h.s. the terms $\propto e^{-i\varphi}$ only. In particular, such terms arise from the contribution $F_{R,+}^{(1)}F_{R,-}^{(1)} 
\partial_y \tilde F_{R,x}^{(1)}$ yielding
\beqa
-D\partial^2_yf_x^{(3)} +2\omega f_x^{(3)}
=
-\frac{2Dq}{\xi_F}\frac{\Delta^3}{\omega^3} a_x
e^{2(y-d_R)/\xi_F},
\label{eq:Usadel4}
\eeqa
where we used that $\xi_F\ll\xi_N$.
A general solution that satisfies \refE{eq:Usadel4} together with the boundary condition $f_x^{(3)}(d_R)=0$ at the right F/S interface reads
\begin{eqnarray}
f_x^{(3)}
&=&\frac{p_x\xi_F}2\frac{\Delta^3}{\omega^3} a_x
\left\{e^{2(y-d_R)/\xi_F}-\cosh p_x(d_R-y)\right\}\nonumber\\
&&+C\sinh p_x(d_R-y).
\label{eq:fx3}
\end{eqnarray}
Here $C$ is a constant which should be determined from the boundary condition at the interface between the left and right ferromagnet. It turns out, however, that it does not contribute to the current and, thus, will not need to be determined. We obtain
\beq
I_2=4\pi e\nu S D T\sum_{\omega>0}\frac{\Delta^4}{\omega^4}p_x^2 \xi_F a_x^2,
\label{eq:I2A}
\eeq
which evaluates to  
\beq
I_{2}=
\frac {\pi G \Delta^4}{192 e T_c^3}\frac{\xi_F}{d_R} \Upsilon_2\!\left(\frac{d_L}{\xi_F},\theta\right)\times\begin{cases}1, &d_R\!\ll\!\xi_N,\\ \frac{384 d_R^2}{\pi^4\xi_N^2} e^{-2d_R/\xi_N}\!,\!\!\!&d_R\!\gg\!\xi_N.\end{cases}
\label{eq:I-2ndharmonic}
\eeq
The second harmonic is long-ranged: it depends on $d_R/\xi_F$ as a power law. The suppression factor $\xi_F/d_R$ is due to the conversion of parallel-spin triplet pairs into singlet Cooper pairs which takes place on a  distance $\xi_F$ from the F/S interface only. The dependence of the function $\Upsilon_2$ that appears in \refE{eq:I-2ndharmonic} on the tilt angle and the thickness of the short layer is shown in Fig.~\ref{2nd-harmonic}. $\Upsilon_2$ is maximal for lengths $d_L \sim \xi_F$ and angles $\theta \sim \pi/2$. As expected, it vanishes for collinear structures, $\theta\to0,\pi$ of arbitrary length. In non-collinear structures, it simplifies to $\Upsilon_2\simeq(d_L/\xi_F)^4\sin^2\theta$ for a short F layer, $d_L\ll\xi_F$, while it vanishes exponentially
\beq
\Upsilon_2=
\frac{4\sin^2\theta\left(\cos\frac{d_L}{\xi_F}\cos^2\frac\theta 2- \sin\frac{d_L}{\xi_F}\right)^2e^{-2\frac{d_L}{\xi_F}}}
{\left(\sin^2\theta+a_\theta\frac{\xi_F}{d_L}\right)^2}
\eeq
for a long F layer, $d_L\gg \xi_F$. Here $a_\theta$ is only relevant in a vicinity $\delta \theta$ near $\theta=0,\pi$, where $a_0=2$ and $a_\pi=1$, respectively.

As in non-collinear structures, the second harmonic $I_2$ is long-ranged (in contrast to the first harmonic $I_1$), we expect it to dominate as soon as $d_R$ exceeds a few times $\xi_F$. In particular, for an optimal thickness $d_L\sim \xi_F$, we find the ratio $I_2/I_1\sim  (\Delta/{T_c} )^2({\xi_F}/{d_R})^2e^{d_R/\xi_F}$ close to $T_c$. The effect is expected to be robust at lower temperatures, further enhancing the ratio $I_2/I_1$ when $\Delta\sim T_c$. More generally, odd and even harmonics will be short-ranged and long-ranged, respectively. However, as triplet pairs need to recombine into singlet Cooper pairs on a distance $\xi_F$ from the F/S interface, we conjecture that the amplitude $I_{2n}$ of the even harmonics $\propto\sin(2n\varphi)$ will contain a small factor $(\xi_F/d_R)^n$, see \refE{eq:I-2ndharmonic} for the case $n=1$. Thus, the current-phase relation could be dominated by the second harmonic at all temperatures.

Let us now discuss the measurability of our prediction. The long-range proximity effect has been observed in trilayers \cite{robinson,birge}, where already the first harmonic is long-ranged. The amplitude of the second harmonic for bilayers predicted in our work is smaller only by a factor $\xi_F/d_R$ as compared to the first harmonic in trilayers. Thus we believe that it should be well within the sensitivity of present-day experiments. Its specific phase dependence may be detected by a direct measurement of the current-phase relation \cite{frolov} or through the appearance of fractional Shapiro steps in the current-voltage characteristics under microwave irradiation \cite{sellier}.

In conclusion, we predicted that the current-phase Josephson relation through a long diffusive ferromagnetic bilayer with non-collinear magnetizations is dominated by a superharmonic contribution $\propto \sin(2\varphi)$. The second harmonic can be viewed as the minimal Josephson current \cite{pals} that can flow between the conventional {\it even}-frequency superconductor in one lead and the effectively {\it odd}-frequency superconductor \cite{bergeret-rev,berezinskii} generated by the long-range triplet proximity effect at the extremity of the ferromagnetic bilayer attached to the other lead. Measuring the dependence of the Josephson current on the thicknesses of the ferromagnetic layers and the angle between their magnetizations would provide further evidence for the long-range triplet proximity effect. Similar experimental studies have been performed for the critical temperature of a thin superconducting film in contact with a ferromagnetic spin valve \cite{leksin}. Furthermore, detecting the $\pi$-periodicity of the current-phase relation through phase-sensitive measurements would be a strong indication of the odd-frequency nature of the long-range proximity effect.
 
\acknowledgements
Part of this research was supported through ANR grants ANR-11-JS04-003-01 and ANR-12-BS04-0016-03, and an EU-FP7 Marie Curie IRG.


\begin{thebibliography}{99}
\bibitem{meso}
In principle, mesoscopic fluctuations may induce a long-range Josephson current even for collinear ordering \cite{spivak,zyuzin}. However, the effect is negligible in metallic junctions.
\bibitem{foot1}
The minimal absolute value of the Matsubara frequency $\omega$ is $|\omega_{\rm min}|=\pi T_c$.
\bibitem{foot2} 
Note that the experiments were performed with single-layer junctions, corresponding to $\theta=0$ in our calculation.
\end{thebibliography}
\end{document}